\documentclass[10pt]{article}
\usepackage[cp1251]{inputenc}
\newcommand{\sss}{\scriptstyle}
\usepackage[dvips]{graphicx}
\def\lsim{\
  \lower-1.2pt\vbox{\hbox{\rlap{$<$}\lower5pt\vbox{\hbox{$\sim$}}}}\ }
\def\gsim{\
  \lower-1.2pt\vbox{\hbox{\rlap{$>$}\lower5pt\vbox{\hbox{$\sim$}}}}\ }

\begin{document}
\title{Uneven horizon \\
      or several words$^*$ about the superfluid $^{4}He$ theory}
 \author{ Maksim D. Tomchenko$^{\diamond}$
 \bigskip \\ {\small   Department of Astrophysics and Elementary Particle Physics,} \\
   {\small Bogoliubov Institute for Theoretical Physics, Kiev 03143,
    Ukraine.}}
  \date{\empty}
 \maketitle
 \large
 \sloppy
   \textit{The state of the superfluid He-II theory is briefly surveyed --- some aspects of its history, achievements, and
      unsolved problems. }\\
      Key Words: Superfluid $^4$He;  Condensate; Quasiparticle Spectrum.

%        \titulstor
 %       \refstor   \nabla   \textbf{\nabla}
 %      "Дорогу жизни пройдя наполовину, \\
 %      вошёл я в тёмный лес..."           \\
 %      Данте, "Божественная комедия"   \\
 %      "Я из лесу вышел,
 %      был сильный мороз..."
 %      Некрасов, "Крестьянские дети"

     \section{Introduction}

    In the first years after the discovery of superfluid He-4,
    the property of superfluidity attracted attention
    not only due to its uncommonness but also to the hope of some physicists for its connection
    with the basically new, still undiscovered laws of the microworld \cite{kap2}.
    After seven decades from the discovery of the superfluidity (P. Kapitsa \cite{kap1}, J. Allen, A. Misener \cite{al-m})
    and a century from the time when He-$4$ was fluidized
    (H. Kamerlingh-Onnes \cite{kam}),
    it becomes clear that superfluid helium justified hopes for the manifestation of a number of unusual properties,
     but no ``new physics'' has arisen because of helium.

       In what follows, I will characterize briefly the state of the theory of superfluid He-4 (He-II) and mark the unsolved questions.
       In this case, I do not pretend to the completeness of analysis or references
       and do not try to avoid the own sympathies for some
       questions. The other reviews on the theory
       of superfluidity He-II are, for example,
       \cite{gl87}--\cite{gl93}, and a short one is presented in \cite{ujp05}.

         \section{History}
          A phenomenological theory of superfluidity was constructed by
          L. Landau \cite{land1,land2} in the 1940s.
          His key idea consisted in that He-II is a quantum
          fluid, in which motions can be only collective,
          i.e., in the form of quasiparticles. L. Landau
          formulated a criterion, according to which a quantum
          fluid is superfluid (SF), if its spectrum of quasiparticles (the curve of the energy $\epsilon$ of quasiparticles as a function
          of momentum $p$) does not touch with the momentum axis, i.e., it has no points with $\epsilon/p \rightarrow 0$. L. Landau proposed
          to describe a gas of quasiparticles as the
         ``normal component'' of the fluid, and the ``rest fluid'' represents the SF-component.
         This yields the two-fluid hydrodynamics of He-II which allowed one to explain a number of properties of helium \cite{pat,xal}.
         But this theory is phenomenological and does not clarify the structure of quasiparticles.
            Therefore, the analysis within the two-fluid hydrodynamics can lead to the erroneous look at phenomena in some cases.

        A microscopic theory of He-II was started in a number of works. F. London assumed \cite{lond} that the superfluidity of helium
        can be related to the presence of a condensate in it. On the basis of this idea, L. Tisza \cite{tissa} constructed
        a two-fluid hydrodynamics of He-II, in which
        the SF-component coincides with the condensate. The last assumption turned out
        false; a more correct two-fluid model
        was developed by L. Landau \cite{land1,land2}. A. Bijl \cite{bijl} proposed an
        approximate structure of the wave  functions of the ground
        and first excited states of He-II:
        \begin{eqnarray}
        \Psi_{0} &\approx &
        Ce^{\lambda\varphi_{1}+\lambda^2\varphi_{2}+\ldots }, \quad
            \varphi_{1}= -\sum\limits_{j>l}\gamma(r_{jl}) -
         \sum\limits_{j}\mu(\textbf{r}_{j}), \label{0} \\
          \Psi_{\textbf{k}}&\approx &
        \Psi_{0}\sum\limits_{j}e^{ikx_{j}},     \nonumber
        \end{eqnarray}
          $\lambda$ --- the interaction parameter.
          The microtheory obtained the strongest stimulus due to the
        works by N. Bogoliubov \cite{bog,bz} and R. Feynman \cite{fey1,fc}.
          In particular, the field-theoretic (FT) approach \cite{bog}
         allowed one to calculate the spectrum of quasiparticles (phonons)
        and the relative number of atoms of a condensate (below --- ``value of a condensate'')
         for the system of interacting bosons under condition for the interaction strength to be small.
        The FT-approach involves the operator formalism and relevant methods. Then it was developed in a number of works,
           in particular, in \cite{bel}--\cite{pash}.

          More significant results were obtained within quantum mechanical (QM) approaches allowing one to calculate, in various ways,
            the full $N$-particle wave functions (WFs) of He-II such as those of the ground state
            ($\Psi_{0}$), the states with one quasiparticle $\Psi_{\textbf{k}}=\psi_{\textbf{k}}\Psi_{0}$,
            and the states with many quasiparticles
             \begin{equation}
    \Psi(\textbf{r}_1,\ldots ,\textbf{r}_N) =
  \prod\limits^n_{i=1}\left (\psi_{\textbf{k}_{i}}
  \right )^{n_{i}}\Psi_0.
  \label{1}     \end{equation}
    It is worth noting that the interaction of quasiparticles in the structure was not considered (\ref{1}).
      The first sequential works in the QM-approach were executed by R. Feynman \cite{fey1,fc}. By using the variational method,
      he obtained the approximate formula
       \begin{equation} E(k)\approx \frac{\hbar^{2}k^{2}}{2mS(k)}
     \label{1f}     \end{equation}
        for the spectrum of quasiparticles (SQ) of He-II, $E(k)$,
      which indicated, for the first time,
      that the roton minimum of $E(k)$
      is related to a maximum of the structure factor $S(k)$, i.e., to a
      short-range order in the fluid.

       The first works by R. Feynman were followed by the articles by N. Bogoliubov,
       D. Zubarev \cite{bz} and R. Jastrow \cite{jast} published simultaneously. In work \cite{bijl}, the logarithm of $\Psi_{0}$
was sought in the form
       of the expansion of (\ref{0}) in the interaction parameter.
               It was shown in \cite{bz} that such a theory of perturbations is not fully correct. There, it was refined, and  a new model of weakly nonideal Bose-gas
        combining the QM- and FT-approaches was proposed; the zero approximation $\psi_{\textbf{k}}= \rho_{-\textbf{k}}$
        was deduced, and the SQ of a Bose-gas,
          \begin{equation} E(k) = \sqrt{\left (\frac{\hbar^{2} k^2}{2m}\right )^{2} +
  2n\nu(k)\frac{\hbar^2 k^2}{2m}}, \quad \nu(k) = \int
  V(r)e^{-i\textbf{k}\textbf{r}}d\textbf{r},
     \label{1b}     \end{equation}
       coinciding with (\ref{1f}) at the weak interaction was found.
       Work \cite{bz} induced the development of QM-approaches with the use
        of collective variables $ \rho_{\textbf{k}} = \frac{1}{\sqrt{N}}
 \sum\limits_{j=1}^{N}e^{-i\textbf{k}\textbf{r}_j} \quad (\textbf{r}_j$ ---
  coordinates of helium atoms). However, that work did not else show that the logarithm of $\Psi_{0}$ can be most explicitly expanded in the correlation series
     \begin{equation}
 \ln{\Psi_{0}}  =  \sum\limits_{j<k}u_{2}(\textbf{r}_{j}-\textbf{r}_{k}) +
    \sum\limits_{j<k<l}u_{3}(\textbf{r}_{j}-\textbf{r}_{k},\textbf{r}_{l}-\textbf{r}_{k},\textbf{r}_{j}-\textbf{r}_{l})
    +\ldots
    \label{4}     \end{equation}
   Similar, but yet not fully correct, structure of WF $\Psi_{0}$ was proposed in \cite{jast},
   where $u_{3}$ was sought in reduced form $u_{3}(\textbf{r}_{j},\textbf{r}_{k},\textbf{r}_{l})\sim f(r_{kj})f(r_{jl})f(r_{kl})$
   and  was not clearly indicated, that just the logarithm of $\Psi_{0}$ should be expanded (it was shown in \cite{bijl,bz}).
    R. Jastrow did not know, probably, work \cite{bijl}, where structure (\ref{4}) was almost deduced.
     The proper structure (\ref{4}) was proposed independently and almost simultaneously by E. Feenberg \cite{feenb0,feenb},
      C.-W. Woo \cite{woo}, T. Regge \cite{redje},
      and I. Vakarchuk, I. Yukhnovskii \cite{yuv0}.
      Moreover, works \cite{yuv0} present a method of calculation of the fine structure of $\Psi_{0}$ and $\Psi_{\textbf{k}}$ [see formulas (\ref{2}) and (\ref{3}) below].

   From the historical viewpoint, it is interesting to note that
      work \cite{bog} was reported by
      N. Bogoliubov at the Landau's seminar, where
      it was criticized. One of the results of that work was SQ (\ref{1b}) which has a minimum in the region of finite momenta
      at a certain potential $V(r)$. L. Landau noticed, without any doubts, this fact
       and submitted his famous work \cite{land2} for publication already in a week,
       where he predicted correctly the spectrum of quasiparticles of He-II measured later on (Fig. 1).
        However, the work by Landau contained no mention about the Bogoliubov's model ...

         \begin{figure}[h]
 % \vspace*{7.5cm}
\centerline{\includegraphics[width=85mm]{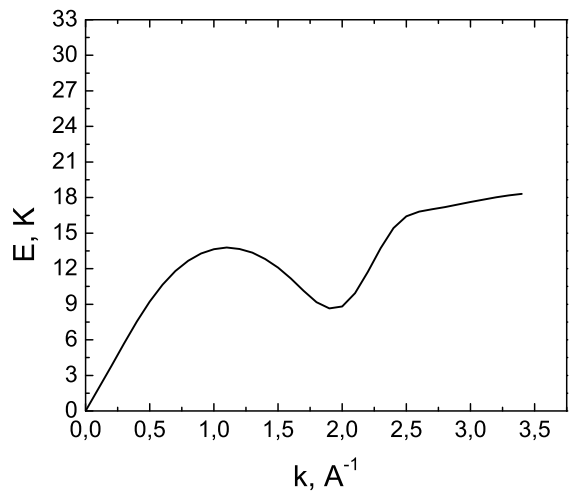}}
\caption{\large \ Experimental spectrum of quasiparticles of
He-II.}
\end{figure}

     I also note that the Soviet researchers in the USSR met difficulties as for the publication on the western journals published in English.
     The direct publications in the West were forbidden by the internal censorship,
     and they were possible mainly with delay through several journals published in Moscow and translated abroad.
      However, due to the hostility between the Bogoliubov's and Landau's schools, the representatives of the former
     had no way to the world scientific literature,
     since this way passed through Moscow journals.
      For this reason, the works by I. Yukhnovskii and I.
      Vakarchuk \cite{yuv0}--\cite{vak}, in which one of the best modern
      QM-methods (see below) on the basis of work \cite{bz} was developed, remained little known.
      One would expect that the period of totalitarian regimes terminates,
      all journals will be open for the authors from any country, and science will be aimed at only constructive purposes.

   \section{Modern quantum-mechanical models of the microstructure of He-II}

  The development of ideas advanced by London, Bijl, and mostly by Feynman and Bogoliubov,
   has led to the appearance of a lot of articles, from which
  four QM-methods should be selected: the methods of collective variables (CV) \cite{redje}--\cite{cond1,ujp05},
     ``correlated basis functions'' (CBF) \cite{cbf-jf}--\cite{cbf-man},
    ``hypernetted chains''  (HNC) \cite{camp}--\cite{krot2d}, and  ``shadow wave functions''  (SWF)
    \cite{vit}--\cite{swf}.

    The method of CV was developed in \cite{yuv}. There, the WFs of the ground and first excited states,
    $\Psi_{0}$ and $\psi_{\textbf{k}}$,
    are sought in the representation of the collective variables $\rho_{{\bf k}}$ in the form of
    infinite series
      \begin{equation}
  \Psi_0 = \frac{e^{{\small S_{0}}}}{\sqrt{Q_{n}}}, \quad
   S_{0} =\sum\limits_{l\geq 2}\frac{N^{1-l/2}}{l!}
   \sum\limits_{{\bf k}_{1},\ldots,{\bf k}_{l}\not= 0}
  \delta({\bf k}_{1}+\ldots+{\bf k}_{l})a_{l}({\bf k}_{1},\ldots,{\bf k}_{l})
   \rho_{{\bf k_{1}}}\ldots\rho_{{\bf k_{l}}},
    \label{2}     \end{equation}
     \begin{equation}
 \psi_{\textbf{k}}  =  \rho_{-\textbf{k}} +
 \sum\limits_{\textbf{k}_{1}\neq 0}^{\textbf{k}_{1} \neq \textbf{k}}
  \frac{b_{2}(\textbf{k},\textbf{k}_{1})}{2!\sqrt{N}}
 \rho_{\textbf{k}_{1} - \textbf{k}}\rho_{-\textbf{k}_{1}} +
  \sum\limits_{\textbf{k}_{1},\textbf{k}_{2}\neq 0}^{\textbf{k}_{1}+\textbf{k}_{2}\not= \textbf{k}}
  \frac{b_{3}(\textbf{k},\textbf{k}_{1},\textbf{k}_{2})}{3!N}
 \rho_{\textbf{k}_{1} + \textbf{k}_{2} -
 \textbf{k}}\rho_{-\textbf{k}_{1}}\rho_{-\textbf{k}_{2}}+\ldots ,
       \label{3} \end{equation}
  where the $l$-th term describes the $l$-particle
   correlations; $\delta$ --- Kronecker delta, $N$ --- the total number of
 helium atoms. By performing the direct substitution of $\Psi_{0}$ (\ref{2}) and
   $\psi_{\textbf{k}}$  (\ref{3}) to the exact $N$-particle Schr\"{o}dinger equation,
   we can verify that $\Psi_{0}$ and
   $\psi_{\textbf{k}}$ satisfy it and can find an infinite chain
   of linked equations for the functions $a_{l}$ and $b_{l}$. These equations are exact,
   but the whole chain cannot be solved analytically or numerically.
   Therefore, it is truncated, and only the first terms describing two-, three-, and sometimes four-particle
   correlations are retained. Such equations can be already
   solved numerically \cite{ujp05,cond1,jetp1}. The method allows one to reproduce the
   experimental SQ of He-II with good accuracy, as well as the energy of the ground state ($E_{0}$) and the values of condensates, but with worse accuracy.
    The analysis can be carried out without fitting
   parameters (FPs), by using the structure factor $S(k)$, or with
   FPs, by using the pairwise potential $V(r)$.

       In the CBF method, the solution for $\psi_{\textbf{k}}$ is also
       sought in the form (\ref{3}), and the functions $b_{l}$ and  $E(k)$ are determined by the Brillouin--Wigner theory of perturbations
       with regard for only, as usual, two first terms in (\ref{3}).
      The method allows one also to calculate the SQ with good accuracy. The best results are obtained in
      \cite{cbf-lee2} without FPs and in \cite{cbf-man} with several FPs.
      In \cite{cbf-jf}--\cite{cbf-lee2}, the WF $\Psi_{0}$ was not calculated. These works used many-particle distribution functions
      under certain assumptions and did not apply FPs.
      In \cite{cbf-man},  $\Psi_{0}$ is sought in the coordinate
      representation (\ref{4}) with regard for two first terms with the functions $u_{2}$ and $u_{3}$
    which are chosen in a definite form with several FPs. In
     CBF models, the four-particle and higher corrections in the expansion of $E(k)$
    were not considered, though taking the subsequent corrections into account changes considerably
    the curve $E(k)$ \cite{cbf-lee2}.

     The HNC models are  developed in \cite{camp,cc1} and
     \cite{krot,krot2d}. In \cite{camp,cc1}, the WF $\Psi_{0}$ is sought in the
     form similar to (\ref{2}), and the functions $a_{l}$
     are determined from the chain of equations obtained by means of variation of the total
     energy. The SQ is calculated by the Brillouin--Wigner theory of perturbations.
     The analysis can be performed without FPs: on the basis of the experimental curve $S(k)$
     or the potential (then $S(k)$ is deduced).
     In \cite{krot,krot2d},  $\Psi_{0}$ is determined in a similar way,
      but in representation (\ref{4}) and with the use of a potential. There, the WF $\psi_{\textbf{k}}$
      and the SQ are calculated within a special method,
      which is equivalent to the calculation of $\psi_{\textbf{k}}$ (\ref{3}).
      % whose essence turns out of our comprehension.
      In HNC models, the quantities $E(k)$, $E_{0}$, $S(k)$, and the value of condensate  can be calculated quite satisfactorily. As drawbacks,
      I indicate that the HNC method itself contains certain approximations, and only two first terms are considered in series (\ref{4})--(\ref{3}).

       Of interest is the SWF method, whose idea consists approximately in the account of the delocalization
       of atoms by the introduction of special simple  ``shadow''  factors in the structure of $\Psi_{0}$ and
   $\Psi_{\textbf{k}}$ in order to partially take higher
   many-particle corrections into account. The quantity $\Psi_{\textbf{k}}$ is sought in the
   form \cite{gal,swf}
         \begin{equation}
 \Psi_{\textbf{k}}   =  \int d\textbf{s}_{1}\ldots d \textbf{s}_{N}\cdot
    \tilde{\sigma}_{\textbf{q}}\cdot \exp{[-C\sum\limits_{j}|\textbf{r}_{j}-\textbf{s}_{j}|^2 -
    \sum\limits_{j<l}(u_{r}(r_{jl})+
    u_{s}(s_{jl}))]},
    \label{5}     \end{equation}
     \begin{equation}
    \tilde{\sigma}_{\textbf{q}} = \sum\limits_{j}\exp{[i\textbf{q}\textbf{s}_{j} + i\textbf{q}A(q)
     \sum\limits_{l(\neq
     j)}(\textbf{s}_{j}-\textbf{s}_{l})\lambda(s_{jl})]},
        \label{6}     \end{equation}
     where $\textbf{s}_{j}$ --- shadow coordinates, $u_{r}(r)=(b/r)^m$ --- the McMillan function,
       $u_{s}(r)=V_{Aziz}(d_{1}\cdot r)/d_{2}$ --- the renormalized Aziz potential, and $C, m, b, d_{1},$ and $d_{2}$ ---
       free parameters. The function $\lambda(s_{jl})$ contains still two FPs; whereas $A(q)$ and, in fact, $u_{s}$ and $u_{r}$ --- free functions.
      The quantity $\Psi_{0}$ is also described by formula (\ref{5}) with $\tilde{\sigma}_{\textbf{q}}=1$.  The structure (\ref{5}), (\ref{6}) was not strictly deduced,
       but it is substantiated by a number of arguments with the use of the analogy with the
       formalism of path integrals.
      At certain values of FPs, it is possible to well reproduce
     the experimental data on $E(k)$, $E_{0}$, $S(k)$, and the value of
   a condensate. A part of FPs is determined under the modeling of $\Psi_{0}$, and the rest of FPs  --- in the calculation of $\Psi_{\textbf{k}}$.
   Note that with  $\tilde{\sigma}_{\textbf{q}} = \sum\limits_{j}\exp{(i\textbf{q}\textbf{s}_{j})}$,
   i.e., without use of  FPs in the derivation of $\Psi_{\textbf{k}}$,  the SQ is reproduced  satisfactorily.
   In \cite{gal}, $\tilde{\sigma}_{\textbf{q}}$ was replaced by
   $\tilde{\sigma}_{\textbf{q}-\textbf{q}^{\prime}}\tilde{\sigma}_{\textbf{q}^{\prime}}$, which improves the agreement for $E(k)$.
   It is probable that the result will be more exact, if $\tilde{\sigma}_{\textbf{q}}$ will be represented by a series of the
    form (\ref{3}) with $\sigma_{\textbf{q}} =
    \sum\limits_{j}\exp{(i\textbf{q}\textbf{s}_{j})}$ instead of
    $\rho_{\textbf{k}}$. In our opinion, the SWF method in such a form does not allow one to significantly improve
    the microtheory of He-II, since
    structure (\ref{5}), (\ref{6})
    involves only partially higher correlations (as the first iteration \cite{vit} of the Schr\"{o}dinger equation
    written in the form of a continual integral), whereas the complete consideration requires to deal with infinite series. In addition,
    the coefficients in (\ref{5}) are not obtained, so that one has to use many free parameters and functions.
    This results in the obscurity whether the agreement with experiment is improved due to a more exact description of the microstructure of helium
    or simply due to a successful selection of a number of FPs.

    Thus, all models take only two- and three-particle corrections into account, by neglecting many-particle ones. In this aspect, the models are equivalent.
    The CV method seems to me the most transparent --- it allows one to obtain the equations
    for the coefficient functions $a_{l}$ and
       $b_{l}$, the required quantities $E(k)$, $E_{0}$, $S(k)$,  and the
   values of condensates exactly and in a comparatively simple way.

   As an efficient numerical tool for the description of the microstructure of He-II, I mention
           the Monte-Carlo method (MC), in which the problem of
          next corrections does not arise, since the full solution of the Schr\"{o}dinger equation is determined numerically at once, though
          only for the ground state $\Psi_{0}$ \cite{gfmc}--\cite{mor}. The MC method gives the most exact description of the ground state,
          its energy $E_{0}$,
          the structure factor $S(k)$, and the values of all condensates. But the method is not also perfect. In particular, it involves a potential $V(r)$
           which is not known exactly for small $r$.
           True, at a high barrier $V_{0}\equiv V(r\rightarrow 0)$, all potentials are
           close to the potential of a hard ball; therefore, the results of calculations for different (but high) barriers depend on the barrier height not so strongly.
           However, some arbitrariness concerning the potential can be noted: e.g., the Aziz potentials include several FPs,
             and one usually \textit{selects} that Aziz potential among several ones which gives the better agreement of theory
            with experiment. Another drawback of the method consists in that
            it is numerical and does not allow one to see
            the analytic structure of a solution and its details. This yields the third drawback:
            due to the error of calculations, the method does not allow one to distinguish two cases where the value of a three-particle (for example) condensate is exactly zero or slightly different from zero,
            say, it equals 0.01.
            From the physical viewpoint, these cases are basically different.
            Moreover, the method does not yield the curve $E(k)$ (and all characteristics of helium for $T>0$), because the
            Schr\"{o}dinger equation involves only the energy $E$, rather than
            the momentum $k$. Therefore, by minimizing  $E$, one determines only $E$ for $k=0$. The whole curve $E(k)$ must be sought
             involving the other methods,
            by choosing, for example, $\psi_{\textbf{k}}$ in the form (\ref{3})
            with variational parameters.

             The literature contains the discussions about the nature of a roton
     \cite{ons,fc,gl93,gal}. A roton is described by WF (\ref{3}) which is the exact solution of the $N$-particle Schr\"{o}dinger equation;
     the same function (\ref{3}) but with small $k$
     describes long-wave phonons. This implies that a roton
     is a phonon, but with short wavelength. To be more exact, a roton is a
     wave packet --- a superposition of infinite in space waves (\ref{3})  with various
     $k$ in the interval ${\sss \triangle} k$. This
     wave packet is compact, with a size of ${\sss \triangle} r \sim 10\,\mbox{\AA}$
     \cite{ion0}, which yields  ${\sss \triangle} k \simeq 1/{\sss \triangle} r \sim 0.1\,\mbox{\AA}^{-1}$.
      That is, a roton is a collective excitation, rather than a one-particle excitation or a
      vortex ring.

            In many works, the potential $V(r)$
          of the pairwise interaction of $He^4$ atoms is considered reliably determined.
          At large distances, $r \gsim 2.6\,\mbox{\AA}$
          \cite{aziz}, the potential is small ($|V|\lsim 10\,K$), negative, and well-known.
           But, at small $r$, the potential is positive, very large ($V_0 \sim  10^2 \div 10^6\,K$), and unknown exactly.
          Indeed, the electron shells of atoms overlap at small $r$, the interaction of electrons and the nuclei
          of both atoms is strong and is modeled \cite{ahlr} in the self-consistent field approximation,
           which gives only an approximate estimate of the potential (with $V_0 \sim   10^6\,K$),
          rather than the exact potential. The barrier $V_0 \sim   10^6\,K$ is also calculated \cite{felt} by approximating the potential
          $V(r \gsim 2\,\mbox{\AA})$ to the region of small $r$ with the use of the quasiclassical approximation, whose exactness is insufficient for the given problem.
           In view of the approximate nature of the construction of the potential at small $r$, it is surprising that many works consider the potential at small $r$
          to be reliably established and to equal the Aziz potential \cite{aziz}.

       On the convergence of QM-methods. In series
       (\ref{4})--(\ref{3}), the corrections  $\sim N$, and there is no small parameter in the equations for $a_{l}$ and $b_{l}$.
       Nevertheless, the solutions for $E(k)$, $E_{0}$, $S(k)$ and the values  of
   condensates should converge to the exact solution with regard for the increasingly greater number
   of correlative corrections in (\ref{4})--(\ref{3}).
      In the chain of linked equations for the functions
         $a_{l}$ \cite{yuv}, each of the equations couples
          the functions $a_{l-1}$, $a_{l}$, and $a_{l+1}$ (and also nonlinear ``lower'' corrections).
          Despite the absence of a small parameter in the problem, the solution for the first functions  ($a_{2}$, $a_{3}$) should converge to the exact one
         if the increasingly greater number of equations in the chain is solved. This
         convergence is related not to a small parameter, but to the consideration of
         higher many-particle correlations.

           What does the theory allow one to calculate? In the frame of QM-methods,
           the SQ can be reliably calculated with regard for only
           two- and three-particle correlations in
           (\ref{4})--(\ref{3}) and without use of FPs \cite{ujp05,cbf-jf}--\cite{cbf-lee2,cc1}.
           The deviation from experimental data $\sim10\,\%$.  The quantity
           $E_{0}$ and the value of a one-particle
           condensate $n_{1}$ are determined with significantly less accuracy, for which the experiment gives $ E_{0}=-7.16\,K$, $n_{1}\approx 0.05\div
           0.1$. These values can be calculated better within the MC method.
           For the condensates, the problem consists in that values of the one- and two-particle condensates at $T=0$ are calculated
         \cite{vak,cond1,cond2} as the exponent of a sum containing
         all $a_{l}$. Since the corresponding
         corrections are not small ($\sim 1/2$), taking the following
          $a_{l}$ into account changes significantly the value of a condensate:
         on the transition from the two- to three-particle correlations, this value varies by several times \cite{cond1,cond2}.
         Thus, if one does not use FPs, then the value of a condensate in the QM-approaches (except the MC method) is calculated with low degree
         of accuracy. To the best of my knowledge, the
         FT-approaches do no contain also a quite correct scheme
         of calculations of the values of condensates for He-II.
           The value of a
           two-particle condensate $n_{2}$ was calculated only in \cite{cond1,cond2,ris}, and no
           reliable experiment for the determination of $n_{2}$ has been proposed till now.
           The structure factor $S(k)$ is determined from
           the interaction potential, and all methods give a good agreement with experimental data.
           As for the potential, the situation looks more complicated. In the region of the potential well,
           $r \gsim 2.6\,\mbox{\AA}$, the models agree with one another, and the potential is satisfactorily
           reconstructed from the structure factor \cite{ujp05}.
           But the height of the potential barrier $V_{0}$ is described by different models differently: from $\sim
           100 \div 1000\,K$ \cite{brueck1,pash,rov,ujp05,jetp1} to $\sim
           10^{6}\,K$ (with the Aziz potential \cite{aziz}; in the frame of the
           MC and SWF methods, such a potential leads to a good
           agreement of theory and experiment for $E_{0}$, $S(k)$ and the
           value of a condensate $n_{1}$). The great difference of $V_{0}$, by 3-4 orders,
           can be related to the efficiency of the potential at
           small $r$, the neglect of many-particle corrections in some models
        \cite{brueck1,pash,rov,ujp05,jetp1} (which can cause the underestimation
        of the potential by 1-2 orders), or the fact that the Aziz potential is overestimated by 1-2 orders due to approximations made in its derivation.

        Let us \textbf{summarize}. The general problem concerning all QM-methods consists in that the exact solution is represented by
        the infinite series (\ref{4})--(\ref{3}), for which
        all terms cannot be calculated. There are several ways to solve the problem:
        1) To realize the break-down in the microtheory, by developing
        a basically new method of solution or the new form of a solution. For example, it is desirable
       ``to convolve'' series (\ref{4})--(\ref{3}) to simpler
       functions which would be exactly determined. But such a method is not discovered
       else. 2) To gradually calculate more and more
       corrections in (\ref{4})--(\ref{3}) without
       FPs or using them only for the potential to a minimal
       extent. The convergence of $E(k)$, $E_{0}$, $S(k)$ and
   the values of condensates can be clearly manifested already after the consideration of, say, 8-particle corrections.
   3) However, a   system of equations involving all corrections up to
   the 8-particle one is  awkward and complicated. At present, in order to treat
   only the 5-particle correction, it is necessary to increase the operating rate and the random-access
   memory of modern personal computers by 3
   orders \cite{jetp1}. It seems that such great efforts are not expedient. Indeed, the studied
   two- and three-particle approximations give a satisfactory agreement
   between theory and experiment, and it is clear that an increase in the number of considered
   corrections will improve the agreement.

   Sometimes, the authors improve the agreement
   with experiment with the help of several
   FPs.  Of course, using
   three and more fitting parameters, it is possible to ``explain'' any simple
   curve such as, e.g., the spectrum of quasiparticles of He-II.  However, many-particle corrections in
   (\ref{4})--(\ref{3}) are not small and, hence, are of importance for the solution. If they will be simply neglected,
   and the curve $E(k)$ will ``be constructed'' with the help of
   several FPs, we will face with a successful fitting, rather than with a description of the real microstructure of He-II. In my opinion, to use
   more than one FP in a model is not expedient.

            The  FT-methods \cite{bog,bel}--\cite{pash}, as far as I know, meet the same difficulty concerning He-II ---
     the input exact equations following from the first principles cannot be solved
     exactly, and one is forced to omit many-particle
     corrections and to solve  ``truncated'' equations.
      In this case, one fails to correctly introduce a small parameter in the problem
      (mainly due to the strong interaction of atoms),
       so that the neglected many-particle corrections are not small, generally speaking.

           I mention also the idea of He-II as a quantum
             crystal \cite{kris1,kris2}.  The authors \cite{kris2} consider that such disordered system as a fluid
              cannot satisfy the requirement for a system to be completely ordered at $T\rightarrow
              0$ (the third principle of thermodynamics). In my opinion,
             helium has no crystalline lattice.
             Otherwise, the SQ would be strongly anisotropic
             like that for the well-known crystalline phases of helium
             \cite{kris3}. In addition, the superfluidity was observed in films
              of helium with thicknesses down to $0.1$  \cite{pl1} and $0.01$  \cite{pl2}
             of the atomic layer at various mean
              distances between atoms, whereas this distance for a crystal must be of a single value equal to the lattice constant.

             \section{Some unsolved problems}

              At the present time, the structure of the composite condensate of helium is not quite clear in that
              how many atoms
              are in the one-, two-particle, and higher
              $s$-particle condensates ($s\geq 3$) and why the structure is just such.
              It is shown in work \cite{cond1} in the three-particle
              approximation that the values of three-particle and higher condensates
               \textit{equal zero} in the ground state of He-II. I note that the three-particle correlations
              were considered, so that it could be expected that the value of the three-particle condensate is nonzero.
           I presuppose that the values of three-particle and higher condensates turn out equal zero with regard for all
           corrections (i.e., for the exact solution) at any temperature. It is possible that the absence of the higher condensates is related, in a way, to
           the fact that the function $a_{l}$ (and $b_{l}$) from (\ref{2}), (\ref{3}) in each equation of the chain \cite{yuv}
   is coupled
   only with two ``adjacent''  $a_{l}$. This pair
   originates from the pairwise potential.

   Above, quasiparticles meant phonons (including rotons). However, in addition to phonons, helium should contain
   also  microscopic
   vortex rings, as a heat excitations, with the radius $r \gsim 3\,\mbox{\AA}$, which was noted by L. Onsager \cite{ons} long ago. Such rings
   were registered in experiments with ions \cite{ion0,ion1,ion2}. The microscopic description of rings
   including the construction of the full $N$-particle WF of a ring
     was not proposed till now. Only a solution in the mean-field approximation
     is available \cite{jons}. In this case, the $N$-particle WF
     of helium with a single ring is sought as the product
     $\prod\limits_{j=1}^{N}\Psi(\textbf{r}_{j},t)$ $N$ of identical
     functions $\Psi(\textbf{r},t)$ satisfying the Gross--Pitaevskii equation
     \cite{gross}. The exact parameters of the least ring (its radius, energy, and momentum) are unknown.
     Respectively, the contributions of rings to the heat capacity of He-II and to the $\lambda$-transition are unknown as well.
     The difference of the experimental heat capacity of helium and
     that of the ensemble of phonons  indicates \cite{jetp3} that the contribution of rings to the heat capacity at
     temperatures $T = 0.8 \div 1.2\,K$ does not exceed
     $6\,\%$. However, it can be significantly greater near $T_{\lambda}$.
     It is possible also that the rings appreciably affect the value of
     $T_{\lambda}$. If this is true, one should observe an anomaly on the curve
     $T_{\lambda}$ as a function of the thickness $d$ of a thin film of helium at $d\sim
     d_{c}\sim 6\,\mbox{\AA}$  \cite{vr2} ($d_{c}$ --- the diameter of the least ring), which
     can be verified in experiments.

       The nature of the  $\lambda$-transition is not quite clear as well.
        A number of models were proposed, and some of them well describe the critical indices.
       But those models are different from the physical viewpoint. In particular, some authors believed that the rings
       play an important role in the $\lambda$-transition
       (see references in \cite{vr2}), but the other ones hold the opposite opinion. The question about the
       nature of the $\lambda$-transition is fundamental,
       because it concerns, in fact, the nature of superfluidity.
       It is considered to be well established that the superfluidity is related to
       1) all motions in He-II are collective, and 2) the spectrum
       of quasiparticles satisfies the Landau criterion. However, the superfluidity is already absent at a
       temperature somewhat greater that $T_{\lambda}$. So that, we must answer the questions:
       What was changed in the medium on the microlevel, and why
       was the superfluidity broken?  This is not quite clear. The process running at
       $T_{\lambda}$ is named sometimes the appearance
       of microturbulence. From the mathematical viewpoint, this means apparently
       that 1) structure (\ref{1}) stops to be a
       solution of the $N$-particle Schr\"{o}dinger equation and 2) a long-range order (LRO) in the medium disappears.  The LRO is determined for the $s$-particle density matrix
    $F_{s}$ of He-II  as
\begin{eqnarray}
&& \lim\limits_{\vert {\bf r}_i-{\bf r}_j^{\prime}\vert\rightarrow
  \infty} F_{s}({\bf r}_1,\ldots,{\bf r}_s\vert {\bf r}_1^{\prime},\ldots,{\bf
  r}_s^{\prime}){\large \vert}_{B} =  \tilde{F}_{s}({\bf r}_1,\ldots,{\bf r}_s)
  \tilde{F}^{*}_{s}({\bf r}_1^{\prime},\ldots,{\bf r}_s^{\prime}),
   \label{7} \\
  &&\lim\limits_{\vert {\bf r}_i-{\bf r}_j\vert\rightarrow
  \infty} \vert\tilde{F}_{s}({\bf r}_1,\ldots,{\bf r}_s)\vert^{2}
    = F_{s}(\infty)=\left [F_{1}(\infty)\right ]^s =const>0 \nonumber
       \end{eqnarray}
    (in the 3D case),
   where $i,j=1,\ldots,s$, and condition $B$ means that $\vert {\bf r}_i-{\bf r}_j\vert$ and
   $\vert {\bf r}_i^{\prime}-{\bf r}_j^{\prime}\vert$ are fixed.
    In a three-dimensional space, just the presence of the LRO causes the appearance
    of all condensates \cite{vak,cond1,cond2}.
    However, the things are not so simple with a LRO. Known are the theorems
    which forbid the existence of LRO in the 2D space in an infinite system. However, all systems in the nature are finite in all directions, and
     the values of a condensates and the form of LRO for such systems are unknown at present.

     This problem was considered for periodic boundary conditions (BCs)  \cite{gu1,gu2}
      which are used explicitly or implicitly in all microscopic models of He-II.
     But since the periodic BCs are not realized, obviously, in practice
      for systems consisting of a great number of particles, the problem could be solved more exactly for zero BCs. In
      this case, the minimal momentum of a particle is not zero, but it is equal to
      $2\pi/L_{j}$ ($L_{j}$ --- sizes of the system, $L_{x}, L_{y}, L_{z}$),
      which follows also from the uncertainty relations.
      The condensate is a fine effect, and we may expect that the proper setting of BCs
      is important for its evaluation. Apparently,
      both one- and two-particle condensates exist in finite systems, including  finite 2D systems, and ``are dispersed'' over a number of the lowest levels.
      In the 3D space, the summation over all levels will result, apparently, in the condensate value
      $N_{\textbf{k}=0}$ for an infinite system.

         In 2D films, the superfluidity
         is conserved \cite{pl1,pl2}. If the condensates and LRO
         exist in such films, like the 3D case, we may conclude that both the condensate
         and long-range order are the microscopic
          peculiarities integral to the property of superfluidity. For
          bounded systems, the criterion for LRO (\ref{7})
          must contain $\vert {\bf x}_i-{\bf x}_j\vert\rightarrow
      L_{x,0} < L_{x}$ instead of $\vert {\bf r}_i-{\bf r}_j\vert\rightarrow
      \infty$, and the analogous substitution will hold for $y$ and $z$.

          It was observed recently that He possesses electric properties \cite{r1}--\cite{r3},
          including those without external electromagnetic fields \cite{r1,r2}.
            For the time being, these interesting experiments have no explanation. The electric
            signals \cite{r1,r2} are likely related to
            the ``spontaneous'' polarization of helium which appears under
            certain conditions without external electromagnetic
            fields, as a result of the mutual polarization of helium atoms
             \cite{lt1,dm2}.

      \section{Conclusion}
        I have tried to describe the state of the theory of superfluid He-II. As
        seen, the theory is gradually developed, but many point remain obscure, including the basic
        properties of the system and the very property of superfluidity.
        This presents a wide field for the future studies and,
        of course, surprises.

          The work is performed under the financial support
   of the Division of Physics and Astronomy of the NAS of Ukraine in the frame of a Target program
    of fundamental studies N 0107U000396. \\ \\
     $^{\diamond}$ {\small Electronic address:
     mtomchenko@bitp.kiev.ua}\\
   $^*$ {\small A numerical expression of the term ``several words'' is not strictly
   definite in the literature, and I set it to be
   approximately 5000 words.}

\renewcommand\refname{References}

       \end{document}